\documentclass{article}
\usepackage{graphicx} 
\usepackage{array}
\usepackage{booktabs}
\usepackage{natbib}
\usepackage{hyperref}
\usepackage{xspace}
\usepackage[most]{tcolorbox}
\usepackage{subcaption}

\hypersetup{
    colorlinks=true,       
    linkcolor=darkblue,    
    citecolor=darkblue,    
    filecolor=darkblue,    
    urlcolor=darkblue      
}

\usepackage{xcolor}
\definecolor{darkblue}{RGB}{0, 0, 139}

\newcommand{\hacc}[0]{Human-AI Co-Creativity\xspace}
\newcommand{\quotebox}[3]{
\begin{center}
\begin{tcolorbox}[colback=gray!10!white, colframe=white, 
                  width=0.8\textwidth, 
                  boxrule=0mm, 
                  sharp corners, 
                  boxsep=2mm, 
                  left=2mm, right=2mm, 
                 ]

\begin{center}
\begingroup 
\setlength{\baselineskip}{1.5\baselineskip} 
{\huge \textbf{``}}\large \itshape #1{\huge\textbf{''}}\\[2pt]
\endgroup
\end{center}

\begin{flushright}
---\textbf{#2} \\
\textit{{\footnotesize #3}}
\end{flushright}
\end{tcolorbox}
\end{center}
}

\title{Human-AI Co-Creativity: Exploring Synergies Across Levels of Creative Collaboration}
\author{
    Jennifer Haase\\
    Weizenbaum Institute and Humboldt University\\ Berlin, Germany\\
    \texttt{jennifer.haase@hu-berlin.de} 
    \and
    Sebastian Pokutta\\
    TU Berlin and Zuse Institute Berlin \\ Berlin, Germany\\
    \texttt{pokutta@zib.de}
}
\date{November 2024}

\begin{document}


\maketitle

\section{Introduction}

Integrating generative AI into creative work signifies a profound shift in how humans engage with digital tools to create. We are entering an era where AI systems do more than support human creativity: they actively participate in co-creative processes, which we refer to as \hacc \citep{colton_computational_2012, serbanescu_human-ai_2023}. Some creative tasks can now be fully automated, which becomes evident, for example, with the generative fill function in Photoshop (see also Adobe Firefly), code generation in IT \citep{tian_is_2023}, or character design in video games \citep{janson_special_2023}. These examples demonstrate generative AI's potential to enhance human creativity, which some argue is the current limit of existing generative AI tools (e.g., \citealt{marrone_how_2024}). However, we argue that \hacc has the potential to enhance human creative capabilities through the integration of (generative) AI tools, systems, and agents far beyond what is currently common for (non-enhanced) human creativity. This paradigm shift demands a deeper understanding of these co-creative interactions, associated challenges, and the requirements for (generative) AI augmentation \citep{melville_putting_2023}.

Improving individual human creative skills and performance is one of the cornerstones of creativity research, with various techniques and manipulation methods being tested \citep{haase_creativity_2023, Sio_impact_2024}. As human lives increasingly shift into the digital realm, these techniques are naturally becoming increasingly digital as well \citep{bereczki_technology-enhanced_2021, rafner_creativity_2023}. Generative AI tools bring a whole new level and potential of competence increase \citep{rafner_creativity_2023}, with ``human-like'' communication skills while at the same time offering much improved beyond-human-like knowledge and information processing skills (see, e.g., GPT-4, \citealt{openai_chatgpt-4_2023}); at least in certain respects. As with all forms of digitization, there is a risk of losing skills versus the chance of gaining more efficiency and output quality through digital support \citep{parasuraman_model_2000}. In the context of creative work, the maximum benefit of AI will be derived where its focus is human-centric and is designed to enhance, rather than replace, human creativity \citep{anantrasirichai_artificial_2022}.

\quotebox{It's not a human move. I've never seen a human play this move. So beautiful.}{Fan Hui}{Then-European Champion's commentary on\\ game between AlphaGo against Lee Sedol}

However, the potential for genuine AI creativity emerged much earlier, with a striking example being DeepMind's AlphaGo defeating world champion Lee Sedol in Go in 2016. AlphaGo first learned from historical match data, then honed its skills by playing millions of games against itself as well as against human experts. This event is often regarded as a cornerstone in recognizing AI's creative capabilities, which, in hindsight, turn out not to be merely isolated anomalies but precursors of the broader creative possibilities that AI systems offer. Coincidentally, these human players also significantly improved their own proficiency at Go while training the AlphaGo system; see \citet{metz_sadness_2016} for a detailed account. We consider this a prime example for the  human creative advancement achieved through training and working with AI engines, i.e., the interactions with AI system have a lasting impact on the user in terms of creative improvement, beyond the times of interactions.

Integrating generative AI tools into creative processes presents an opportunity to advance human creative performances collaboratively. By focusing on augmenting rather than replacing human creativity, these tools can help overcome the limitations of traditional methods and push the boundaries of what is creatively possible. In this chapter, we will discuss the evolution of creativity support through digital tools, moving from simple digital aids to partially automated tools, culminating in collaboration between humans and generative AI tools. First, we elaborate on the ``inherent'' creative potential of (generative) AI tools, which we posit to be a requirement for actual co-creativity. Then, we differentiate between different forms of digital tool support. By presenting concrete examples from mathematics for varying levels of human-AI co-creative interactions, we will illustrate how the co-creative process with generative AI can significantly advance the creative outcome, achieving new results often with creative twists beyond previously known approaches and, due to their high irregularity, unlikely to be found by human creativity alone.

\section{Creativity of Generative AI tools}
For a system to be considered autonomously creative, it must possess the potential for creative action, such as generating novel ideas or solutions independently without human intervention \citep{jennings_developing_2010}. This then points to the question of \textit{inherent} creativity of generative AI tools. Machine learning serves as the cornerstone for such a form of creativity, providing the capability for algorithms to learn, adapt, and respond in a manner that can be deemed ``intelligent''---and thus, potentially, creative \citep{mateja_towards_2021}.

However, the debate surrounding the ``true'' creativity of technical systems transcends scientific inquiry and becomes a philosophical debate about \textit{appearing} vs. \textit{being}. This discourse revolves around the potential limitations of generative AI, with some viewpoints suggesting that AI's reliance on pre-existing data would confine it to only displaying ``incremental creativity'', thus questioning the depth and authenticity of its creative output \citep{boden_computer_2009, cropley_creativity_2023}. Particularly in non-scientific literature, there is a prevalent notion that only humans with their unique capacity for emotions and empathy could exhibit true creativity \citep{joshi_can_2022, white_opinion_2023}. This perspective is echoed by \citet{runco_ai_2023}, who suggests that the process of creativity in AI, being fundamentally different from the human approach, can only result in what could be termed ``artificial creativity''. We do not share such notions of diminishing the creative output from artificial agents. As we move from the philosophical to the practical, we can see empirical evidence for significantly increased creativity in (generative) AI tools and agents output and human output in collaboration with generative AI tools. Large language models (LLMs), for example, are specifically designed to balance factual precision with creative expression, incorporating elements of flexibility and randomness that allow generating content perceived as original and inventive \citep{sinha_mathematical_2023}. These models leverage vast datasets and complex algorithms to synthesize information in novel ways, resulting in outputs that emulate human-like creativity and demonstrate the potential for independent creative thought within specific domains \citep{rafner_creativity_2023}.

Empirical studies further support the inherent creativity of AI systems. Standardized creativity tests, traditionally used to measure human creativity, have been adapted to evaluate the outputs of generative AI. The results are striking, with AI-generated content sometimes matching or even exceeding human performance in tasks that measure everyday originality and elaboration \citep{gilhooly_ai_2023, guzik_originality_2023, haase_artificial_2023}. Moreover, AI-generated outputs have proven so convincing in practical scenarios to even fool experts in whether content was created by humans or AI (e.g., with scientific abstract, \citealt{else_abstracts_2023}; with artificially generated art, \citealt{haase_art_2023}), one of the most substantial possible benchmarks. This evidence underscores the argument that generative AI tools possess inherent creativity, characterized by their ability to autonomously produce novel and valuable output and pass the test of being indistinguishable from human output.


\section{From digital tools to AI}



Throughout history, tools have been essential to human creativity. Naturally, since the advent of computers, this creative work has increasingly moved into the digital domain. For example, every text editor enables and supports creative writing. While some tools transfer the creative task into the digital, others are designed to engage more actively in the creative process (cf. Table 1). We categorize such digital tools into four distinct types. The first is a \emph{Digital Pen} akin to creative support systems (CSS), which aid human creativity without directly contributing creative input, just like a painting program provides a digital brush to an artist \citep{shneiderman_creativity_2007}. The second type is \emph{AI Task Specialist}, which is an independent AI system (often a generative one) that operates autonomously without human intervention (apart from the initial input). Examples include non-deterministic algorithms that generate art via generative adversarial neural networks \citep{hitsuwari_does_2023} or algorithms that advance game development \citep{almeida_reinforcement_2023}. The third type is a \emph{Creative Assistant}, a generative AI tool that supports and enhances various aspects of a human-driven creative process, often in an interactive way. Current generations of LLMs, such as, e.g., ChatGPT, Gemini, or Llama, are prime examples of that category. Users can flexibly use such tools to support their brainstorming tasks (e.g., \citealt{fui-hoon_nah_generative_2023}) or concrete problem-solving tasks such as coding (e.g., \citealt{dellaversana_gpt-3_2023}). The fourth level, as most pertinent to this discussion, is co-creative systems, which we dub \emph{AI Co-Creators}. Here, humans and (generative) AI tools collaborate, each contributing to the creative process. Ideally, such a system adapts flexibly to the user's needs, can solve complex, open-ended problems and contributes input in a measurable and meaningful way to the co-creative process with the human user.

\begin{table}[ht]
\footnotesize
\centering
\begin{tabular}[t]{p{1.8cm}p{2.0cm}p{2.0cm}p{2.0cm}p{2.0cm}}
\toprule
\textbf{Level of AI} & \textbf{Level 1:} & \textbf{Level 2:} & \textbf{Level 3:} & \textbf{Level 4:} \\ 
\textbf{integration} & \textbf{Digital Pen} & \textbf{AI Task Specialist} & \textbf{AI Assistant} & \textbf{AI Co-Creator} \\
\midrule
\textbf{Description} & Digital tool that facilitates the conversion of traditional creative processes into digital formats & AI tool that augments creative tasks, operating with structured guidance and user input & Generative AI tool enhances everyday creativity, working within the scope of its training data and user prompts & Generative AI tool that generates original ideas and engages in creative dialogue, adapting within set ethical and creative boundaries  \\ 
\addlinespace[1em]
\textbf{Example} & Classical CSS & Generative Autofill by Adobe Firefly & Current LLMs like GPT-4 or Midjourney & domain-specific examples exist \\ \addlinespace[1em]
\textbf{Tool-contribution} & Digitalizing creative work, improving knowledge transfer and communication & Automation of creativity based on strong guardrails and user prompting & Creative on everyday creativity level, limited to training data; based on user prompting & Equal collaborator, original and useful contribution to a shared creative process; argues with a user; based on meta-calibration and intent within broader guardrails \\ \addlinespace[1em]
\textbf{Breakdown of contribution} & Basic assistance in digitalizing traditional creative content & Moderate augmentation in specific creative tasks & Significant enhancement in shaping the final creative product & Synergistic partnership with equal input on creative outcomes \\ 
\bottomrule 
\end{tabular}
\caption{Four levels of human-tool interaction}
\end{table}
 
The four levels indicate the degree of interaction between the user and the tool, depending on how creatively competent and potentially autonomous the tool can act. To demonstrate the varying levels of AI-human interaction in creative processes, we turn to examples from the field of mathematics. We chose mathematics because it allows for objective evaluation of creativity in terms of newness and usefulness, this is in contrast to ``subjective disciplines'' where a direct attribution of usefulness can sometimes be difficult. Although often perceived as rigid, mathematics is inherently creative, demanding innovative approaches to solve complex problems and develop elegant proofs. The study of creativity itself draws from mathematical insights, as evidenced by \citet{wallas_art_1926}, whose model of the creative process is rooted in earlier work by mathematicians like \citet{poincare_mathematical_1908} and echoed in Hadamard’s later contributions \citeyearpar{hadamard_essay_1954}. 

In the following, we will present the four levels of human-tool interaction, with three examples for levels 2-4 of mathematics demonstrating \hacc on various complexity levels. For Level 1, the Digital Pen, basically every general-purpose collaboration tool, like email, Slack, Discord, or Github, would be an example of how researchers communicate and coordinate their creative work. We deem this rather known and familiar to the reader and, for the sake of brevity, do not provide further examples. For the other examples, we will briefly describe the underlying mathematical problem for the non-expert. We apologize to the expert readers for the simplification here, which is necessary to keep the exposition on point and not to deviate into technical details. Moreover, we focus on the three examples from the second author's research. We stress that this might add a particular anecdotal component to the discussion. Indeed, there is a vast body of work in mathematics using AI systems on various levels to achieve new results. However, it also provides us with a higher degree of introspection into the creative process that is usually unavailable as the focus is on reporting results and not processes.

\subsection{Level 1: Digital Pen}
The first level represents the traditional approach of how information systems have long supported humans in their creative processes, with CSS evolving from simple digital tools to complex systems that offer collaborative support and process guidance \citep{muller-wienbergen_leaving_2011, voigt_improving_2014}. These systems have transitioned from mimicking traditional tools to providing process support by integrating advanced knowledge and communication management features \citep{frich_twenty_2018, voigt_improving_2014}. Such tools digitalize and simplify individual or group processes, support the collection, editing, and visualization of human-generated ideas \citep{olszak_conceptual_2018, voigt_improving_2014} but do not address the essence of the creative process itself. Although effective in facilitating creativity, these systems remain tools rather than active contributors to the creative process.

Only with tools integrating some form of (generative) AI can some degree of inherent creativity be assumed to emerge; otherwise, no such entity can contribute to the creative process. AI has the potential to process information, aggregate knowledge, and generalize beyond its training data with the possibility of exceeding human competencies and capacities. The idea of CSS, being support systems for the idea generation process, has so far only been realized in a relatively weak form. However, with the advent of artificial intelligence, a paradigm shift, similar to what has been observed in other disciplines, is emerging: Machine-learning algorithms in AI systems can create content and, with that, potentially creative output \citep{seidel_artificial_2020}. These content-creation functions can either be used to substitute parts of the originally human-only creative process (Level 2) or support and augment various aspects of the creative process (Level 3). 

\subsection{Level 2: AI Task Specialist}
In Level 2 interactions, the human defines the creative problem by specifying parameters and constraints, while the AI performs complex computations at a scale and speed unattainable by the human alone. The AI serves as a highly efficient tool, extending the human’s creative capacity by executing tasks that would otherwise limit exploration due to their complexity or resource constraints. The human remains the primary source of creative insight, with the AI operating within clearly defined boundaries. This interaction is characterized by a high degree of human control over the creative outcome, with AI functioning as an enhancer of human capabilities.

Advancements in rapid and efficient data processing, as seen in tools like Adobe Firefly, exemplify the capabilities of Level 2 systems. These systems enable quick information generation, such as visual auto-fill functions, where AI can extend or substitute parts of a picture with generated content, allowing the user to iterate faster and explore a broader range of ideas. While such tools demonstrate an inherent, albeit rudimentary, form of creativity by generating new and potentially useful content, their creativity is largely incremental, as described by \cite{cropley_creativity_2023}. The user’s interaction remains limited to a specific creative task, and the AI operates under restricted parameters, offering only partial creative autonomy.

\subsubsection*{Math example: New Bell inequalities}

A central question in quantum physics, particularly quantum mechanics, is to decide whether a given state exhibits \emph{quantum behavior} or is just a classical state in disguise. Strongly related to this question are, for example, the central questions for several of today's quantum computer designs: Are they actually quantum computers or just classical ones in complicated designs? To prove that a state is genuinely non-classical, typically, physicists devise a series of clever measurements that exhibit behavior that cannot be explained with classical physics; there are also ways of proving that a state is classical via so-called local models. This approach and the associated concept of \emph{non-locality} has been central to establishing the existence of quantum effects dating back to the famous work of \citet{bell_einstein_1964} that resolved the Einstein-Podolsky-Rosen paradox by providing measurements (so-called \emph{Bell inequalities}) that proved that the experiment of \citet{einstein_can_1935} exhibits true quantum entanglement and associated quantum effects. However, once the states that need to be analyzed become more complex and might even be in a very low dimension, the required insight into the underlying structure of physics and the necessary creative design of such measurements is tough to achieve. 
In \citet{designolle_improved_2023}, an AI system was devised, predominantly relying on the so-called Frank-Wolfe methods \citep{braun_conditional_2023}, to support the user in his effort to devise new measurement strategies for complex states. Here, to compute new Bell inequalities for previously unstudied states, the human user specifies the state and all other system parameters, and the AI system then performs a large and complex series of computations (typically weeks on high-performance compute clusters) to compute a series of measurements and the associated (new) Bell inequality. The user then verifies this inequality via straightforward calculations. 

All creative input in this example comes from the researcher, with the AI system providing highly specialized computations at extreme speed and scale. The AI augments the user’s creative capabilities by enabling large-scale exploration but does not generate creative output beyond the predefined task specification. \citet{designolle_improved_2023} were able to derive a wide range of new Bell inequalities for many important scenarios. 

\subsection{Level 3: AI Assistant}
Level 3 systems, the development of generative AI tools such as ChatGPT and, more broadly, \emph{General Pretrained Transformers (GPTs)}, stable diffusion models, and others, their general applicability allows users to receive broader and more personalized support for their own creative challenges: GPTs are GPTs (General Purpose Technologies). The current generation of LLMs like GPT-4o, Gemini, Claude, and others are perceived as competent enough to support humans in a wide range of creative tasks (e.g., for coding, \citealt{liu_is_2023}; story writing, \citealt{doshi_generative_2024}; problem-solving \citealt{heyman_supermind_2024}). Here, the level of creativity that can be achieved is human-limited, as the challenge lies in understanding and leveraging the potential of the underlying competencies of the tool (e.g., for ChatGPT, \citealt{cromwell_discovering_2023}). This stresses a significant point: the capabilities of generative AI must be made usable for humans, i.e., it is about interfacing. For example, the breakthrough of GPT version 3.5 from OpenAI, along with its wider acceptance, occurred when an intuitive chat-based conversational front-end was introduced; a form of unhobbling (essentially removing the handbrakes of highly potent models,  \citealt{aschenbrenner_situational_2024}). However, current LLMs are designed with specific data sources and generalization capabilities, which, while robust, are guided by carefully implemented restrictions and guardrails. These measures, though occasionally limiting, are essential to ensuring the responsible and ethical use of AI, ultimately enhancing the safety and reliability of the creative process. In addition, hallucinations of factual wrong content are common for LLMs \citep{jesson_estimating_2024}, which, however, might not be as relevant for the generation of new creative output compared to the more mundane generation of factually correct essays or reports. It might help you become a great artist, but not necessarily in your homework assignment. In fact, hallucinations might even improve their creative potential to some extent.

\subsubsection*{Math example: New Ramsey Multiplicity Bounds}

A central challenge in graph theory (a graph consisting of nodes and edges) 
is to understand how often specific subgraphs, like cliques (``everyone knows everyone'') or independent sets (``no one knows anyone''), can appear within larger graphs. This problem is closely tied to classical questions posed by Erdős, which have driven much of the research in this area. For instance, determining the frequency of cliques of four or five nodes 
in larger structures is crucial for understanding the broader behavior of graphs. Researchers often rely on sophisticated mathematical tools and intricate constructions to tackle these questions. In \citet{parczyk_new_2024}, an AI system was designed to 
resolve a longstanding problem about the minimum number of independent sets of size four in graphs where the largest complete subgraph has at most four nodes. 
The obtained constructions with sizes of around 800 nodes and more are usually beyond what can be achieved with ad-hoc methods. 

The AI system designed for this task in \citet{parczyk_new_2024} employs advanced search heuristics to discover new constructions. Here, the creative potential is already shared between the human and the AI system. While the user specifies the requirements for the type of construction needed, the AI system delivers the actual construction. The correctness of the construction can then be verified by the human. However, the power of the interaction between humans and AI systems goes beyond mere constructions. It also reveals that optimal constructions are stable and repeatable, giving insight into the underlying structure.

\subsection{Level 4: The AI Co-Creator}

At Level 4, \hacc represents a fusion of human creativity with advanced AI capabilities, where both entities contribute significantly to a shared creative product \citep{davis_human-computer_2013}. In such systems, the inputs and outputs of humans and AI blend seamlessly, resulting in a synergistic creative process that transcends traditional boundaries of human or machine creativity. This co-creative dynamic fundamentally alters the nature of the creative process by positioning the AI not merely as a tool but as an active participant—an "equal"—in the creative process. Like traditional co-creativity among humans, effective Human-AI collaboration relies on shared goals, diverse perspectives, and extensive communication, ensuring that the strengths of both human creativity and AI are fully leveraged \citep{paulus_chapter_2012}.

At this level, AI and humans operate in true co-creative synergy. The AI is capable of independently generating creative outputs—such as new, highly non-intuitive solutions—that go beyond the scope of human preconceptions. The human and AI continuously interact, with the AI generating novel solutions based on minimal input and the human refining and integrating these into the broader creative context. In this form of interaction, AI becomes an equal creative partner, contributing original and meaningful input that the human alone may not achieve. This level represents the full realization of Human-AI Co-Creativity, where both entities' contributions are equally essential for creative breakthroughs.

In this co-creative process, the role of human creators is elevated, requiring them to possess not only creative skills but also a deep understanding of how to effectively interact with AI co-creators. Human creators must be adept at framing creative problems in ways that are compatible with AI's strengths, ensuring that the AI's contributions align with the creative goals. Additionally, human creators need to evaluate and refine the partial results generated by the AI, applying principles such as the MAYa principle (Most Advanced Yet accessible), which, in turn, is based on the well-known MAYA principle (Most Advanced Yet Acceptable; see, e.g., \citealt{hekkert_most_2003}), to ensure that the AI's outputs are novel yet accessible to the human user.

The principles of interaction in \hacc are critical to the success of the collaboration. Shneiderman \citeyearpar{shneiderman_bridging_2020} argues that human-centered AI should be designed to support and enhance human activities, including creativity. He proposes several key concepts to guide the development of these systems: First, maintaining a balance between human oversight and automated operations is essential. This ensures that, while AI provides substantial creative contributions, humans retain control over the final output, preserving the integrity of the creative process. Second, AI co-creators should be designed to augment human capabilities, acting as powerful agents that enhance creativity rather than merely mimicking human skills. Thus, at this advanced level of co-creativity, AI becomes a fully integrated creative partner, contributing ideas that would not emerge through human effort alone.


\begin{figure}[h!]
	\centering
	\begin{minipage}[b]{0.45\textwidth}
		\centering
		\includegraphics[width=\textwidth]{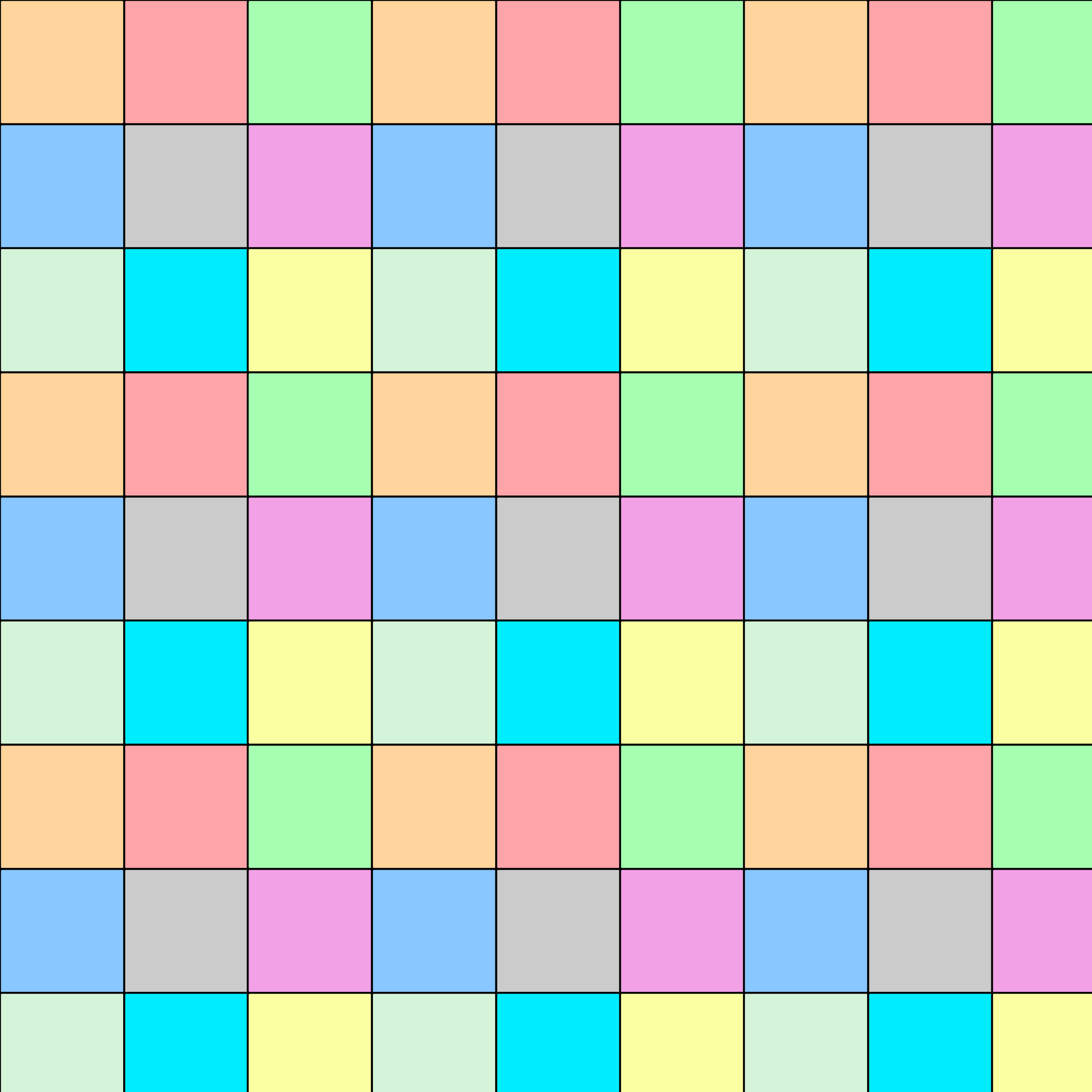}
		\subcaption{$9$-coloring of the plane}
	\end{minipage}
	\hfill
	\begin{minipage}[b]{0.45\textwidth}
		\centering
		\includegraphics[width=\textwidth]{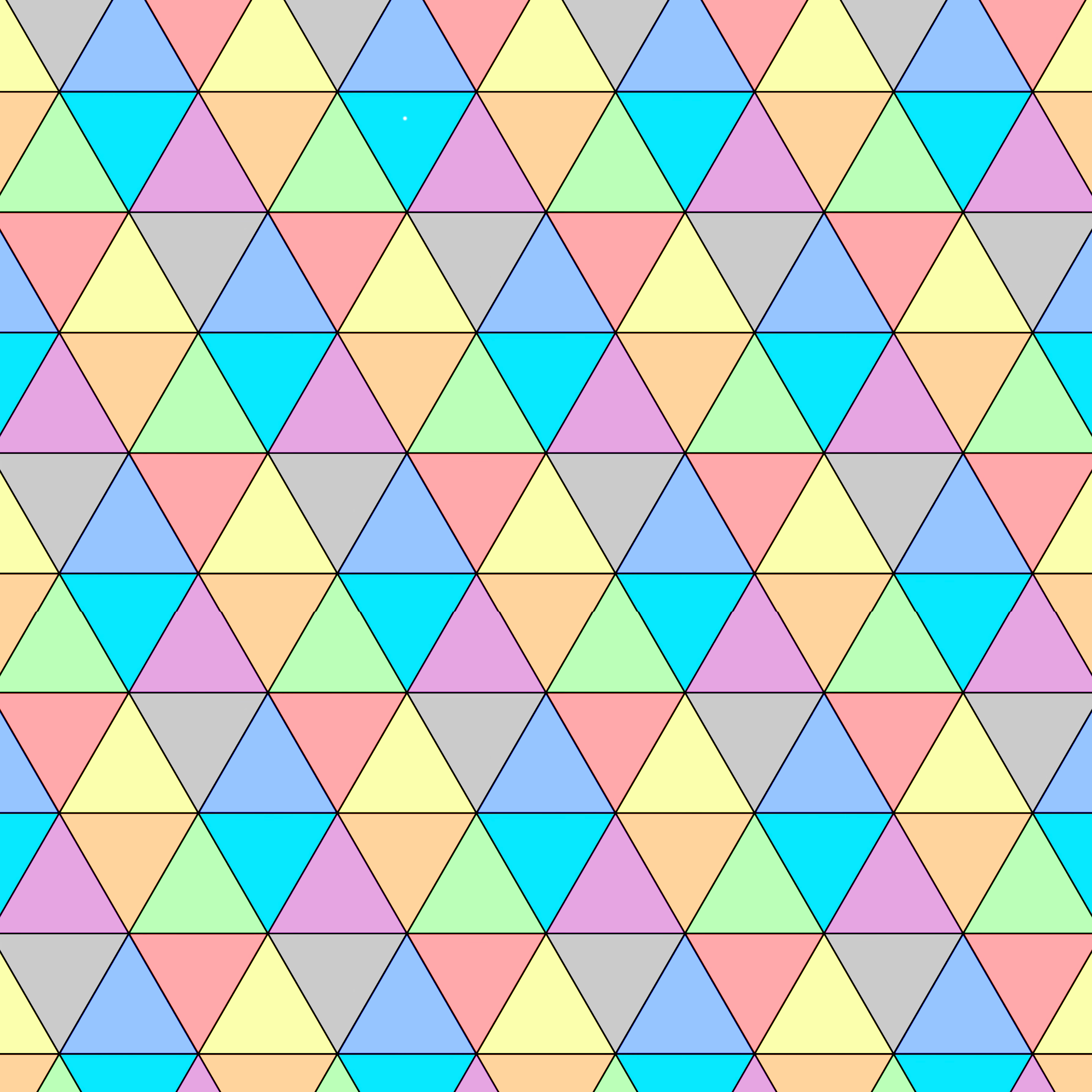}
		\subcaption{$8$-coloring of the plane}
	\end{minipage}
	\vfill
	\begin{minipage}[b]{0.45\textwidth}
		\centering
		\includegraphics[width=\textwidth]{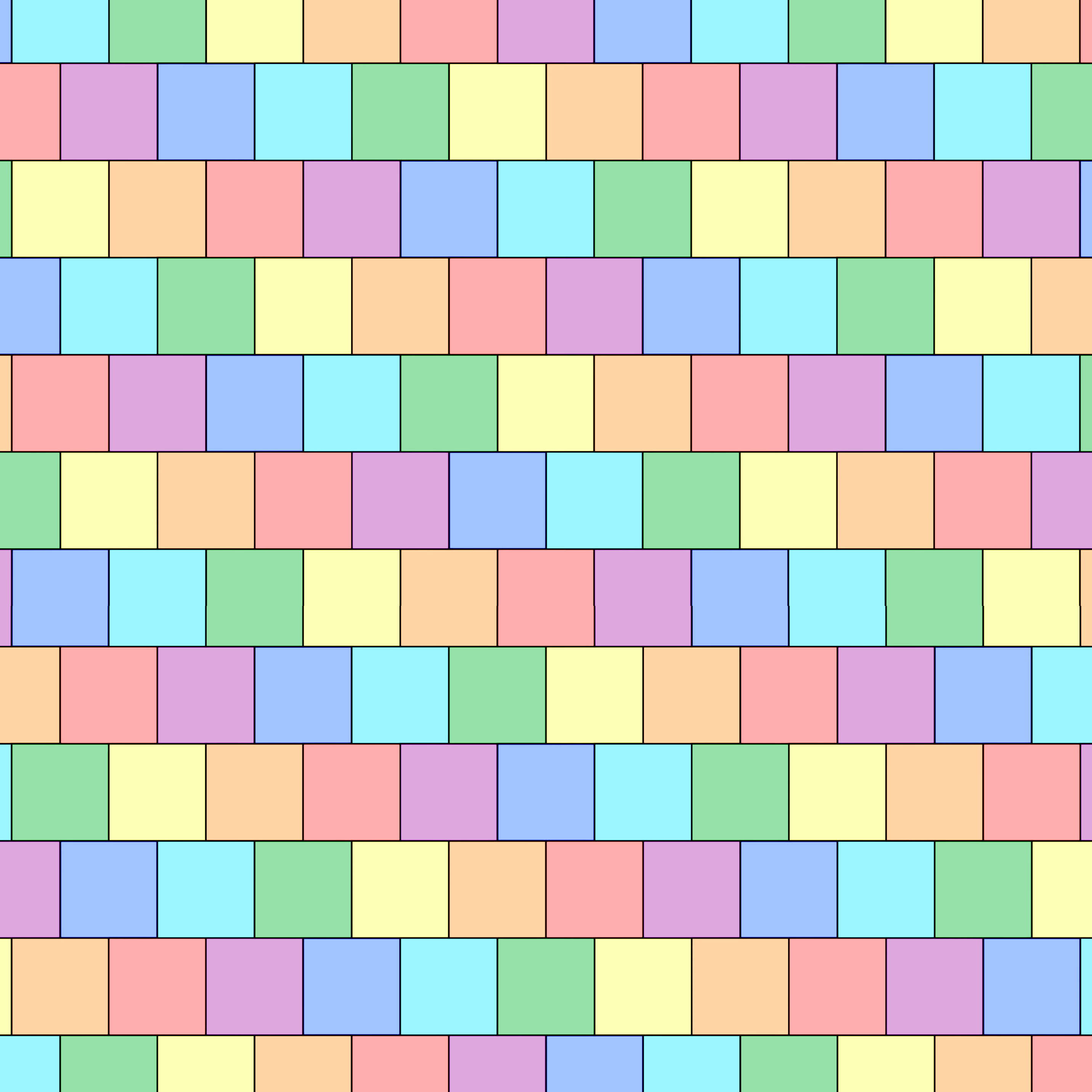}
		\subcaption{$7$-coloring of the plane}
	\end{minipage}
	\hfill
	\begin{minipage}[b]{0.45\textwidth}
		\centering
		\includegraphics[width=\textwidth]{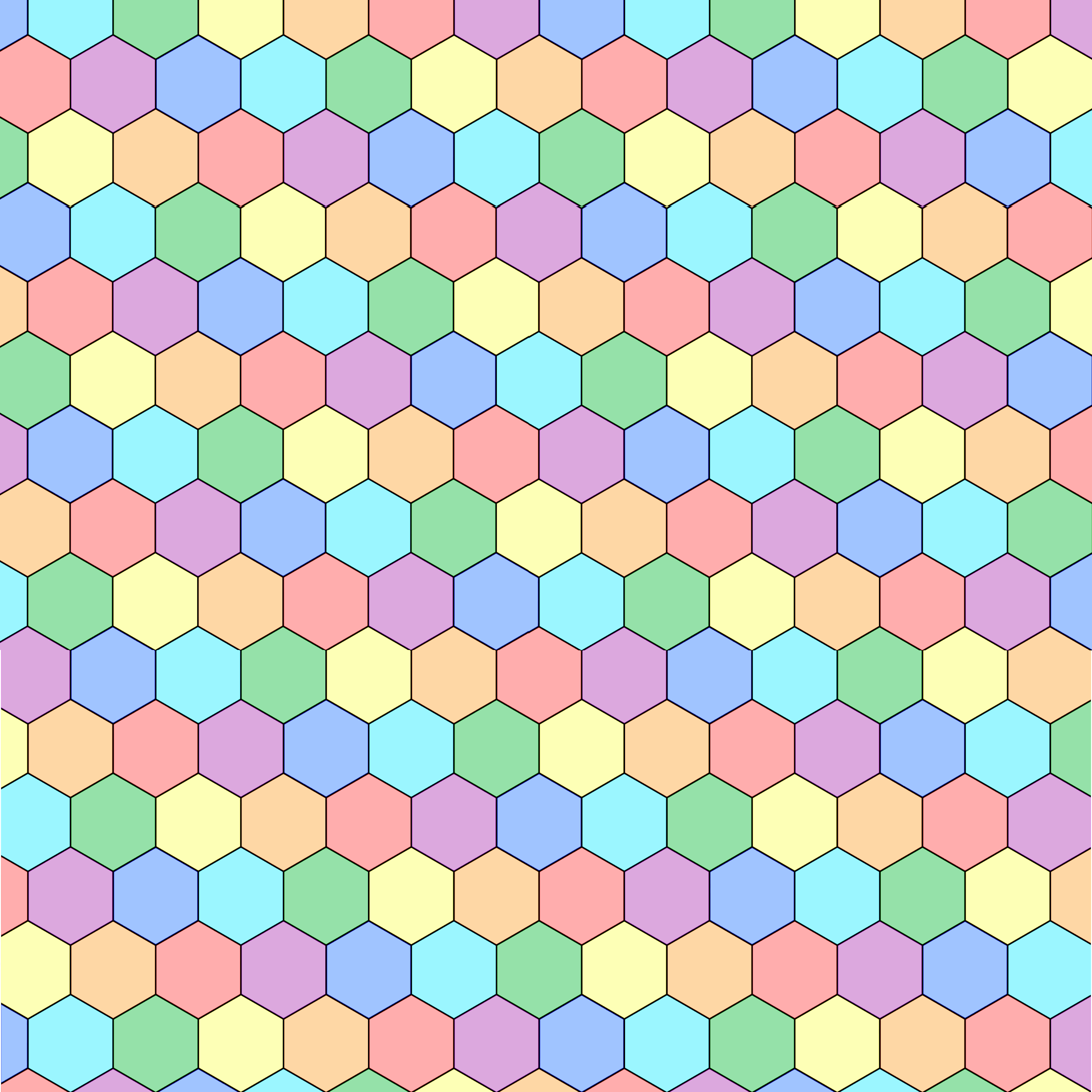}
		\subcaption{$7$-coloring of the plane (alternative)}
	\end{minipage}
	\caption{\label{fig:oldColorings} Known colorings of the plane}
\end{figure}

\subsubsection*{Math example: New Colorings of the Plane}

A central question in combinatorial geometry is the Hadwiger-Nelson problem, which asks for the minimum number of colors required to color the points of a plane so that no two points at a unit distance share the same color. This number, known as the chromatic number of the plane, has intrigued mathematicians for decades; see \citet{soifer_new_2024} for an overview. Recent advancements in this area focus on extending the continuum of valid distances for six colors of the plane. For this purpose, researchers have to construct colorings of the plane with the required properties; see, e.g., Figure~\ref{fig:oldColorings} for a few examples of colorings of the plane. New colorings that go beyond those presented in Figure~\ref{fig:oldColorings} are very hard to find and require a high degree of ingenuity and creativity. There has not been any significant progress for the last $30$ years. Then, in recent work in \citet{mundinger_extending_2024}, two new six-colorings that avoid monochromatic pairs of points at a unit distance for the first five colors and another specified distance $d$ for the sixth color were presented, which were obtained through a customized AI approach.

\begin{figure}[h!]
	\centering
	\begin{minipage}[b]{0.45\textwidth}
		\centering
		\includegraphics[width=\textwidth]{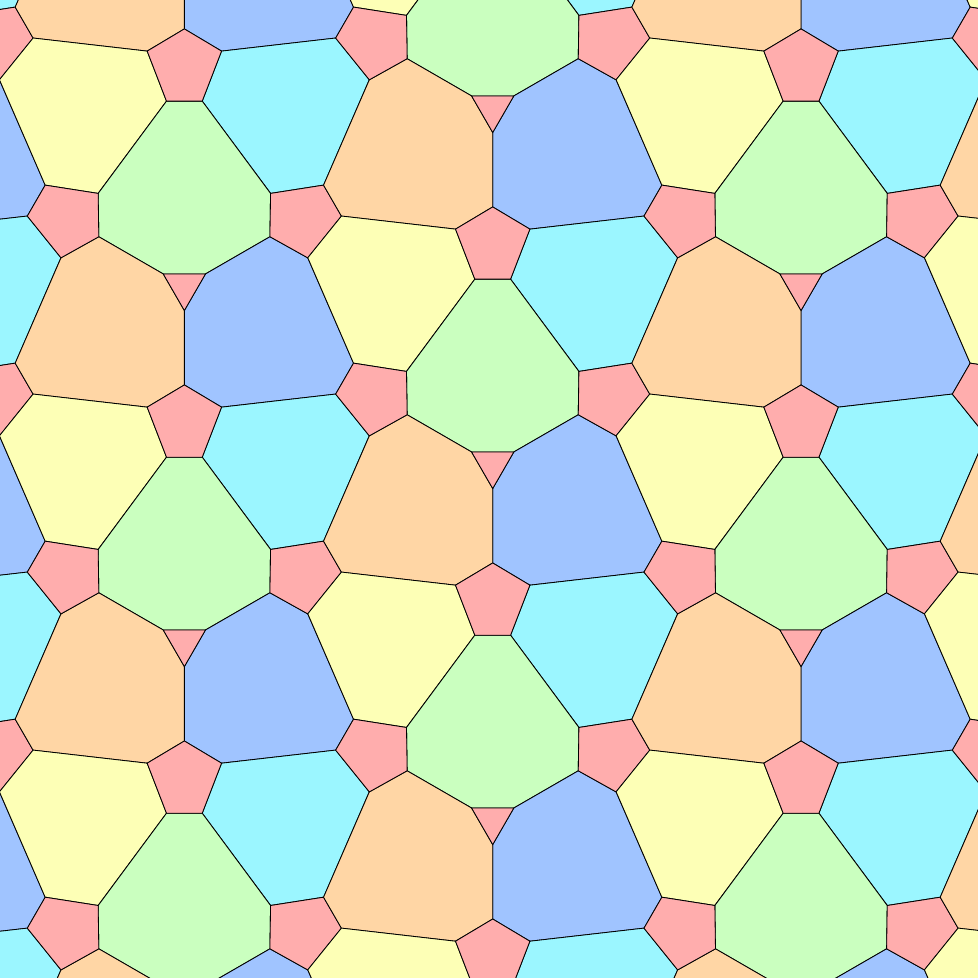}
		\subcaption{\(0.354 \leq d \leq 0.553\)}
	\end{minipage}
	\hfill
	\begin{minipage}[b]{0.45\textwidth}
		\centering
		\includegraphics[width=\textwidth]{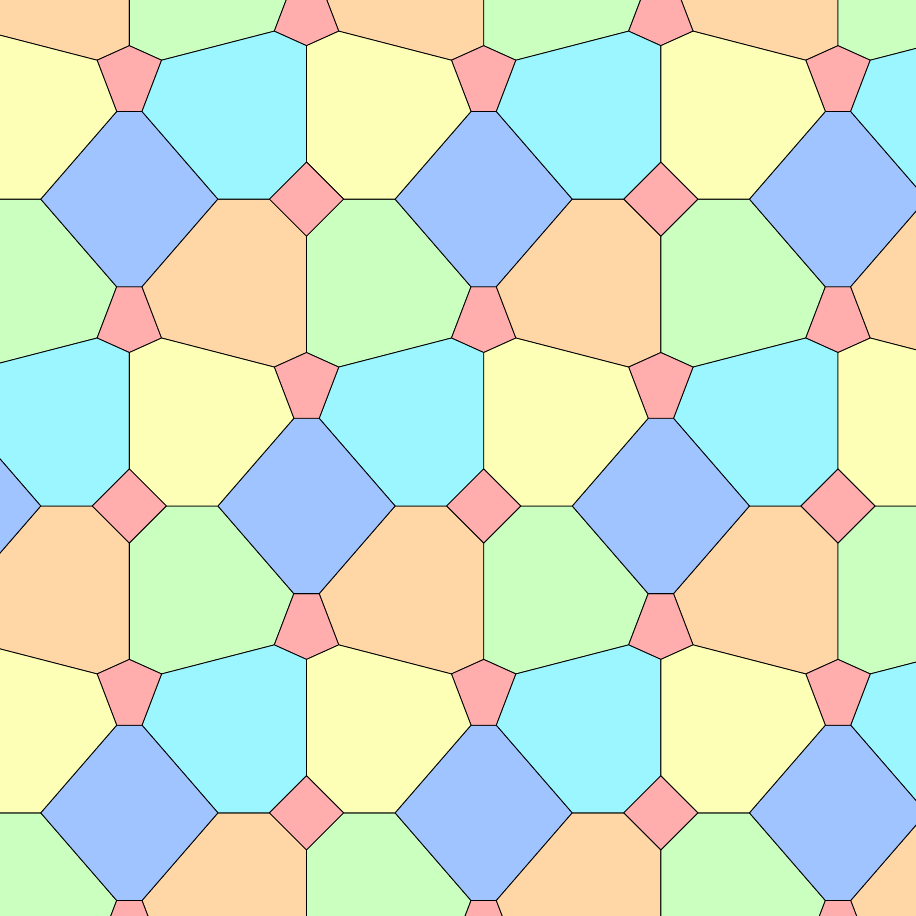}
		\subcaption{\(0.418 \leq d \leq 0.657\)}
	\end{minipage}
	\caption{\label{fig:newColorings} Two new $6$-colorings obtained via \hacc}
\end{figure}

While not entirely a Level 4 system yet, due to its particular purpose, in contrast to the previously mentioned examples, the generative AI system only gets the requirements that a correct coloring needs to satisfy as an input. Then, the system is trained to explore and identify new colorings and construct and evaluate new colorings efficiently. This led to the discovery of the two aforementioned new six colorings satisfying the modified requirement regarding the sixth color, significantly expanding the known range for these colorings. Moreover, the obtained colorings (see Figure~\ref{fig:newColorings}) are highly non-intuitive and creative, breaking the highly symmetric patterns of previous colorings found by humans via trial-and-error, intelligent guessing, and ad-hoc approaches (cf. Figure~\ref{fig:oldColorings}). As before and customary in mathematics, the obtained colorings were then verified and post-processed by a human.

\section{Discussion}

The implications of AI in creative work are multifaceted and far-reaching. As Cremer et al. \citeyear{cremer_how_2023} outline, AI might take several plausible paths to disrupt creative work. Firstly, AI could lead to an explosion of AI-assisted innovation, enhancing human creativity without necessarily replacing it. This democratization of innovation is exemplified by tools like GitHub’s Copilot, which aids in coding by providing real-time suggestions that augment human efforts \citep{cambon_early_2023, eapen_how_2023}. Secondly, there is the potential for AI to monopolize creativity in specific fields, such as game design, where AI-generated art increasingly replaces human designers \citep{christofferson_how_2023}. Lastly, a scenario may emerge where “human-made” creativity commands a premium, preserving a competitive edge over AI-generated content. This preference for human involvement has been noted in experiments where human-generated works were received more positively when a human label was added than when they were tagged with an AI label \citep{bellaiche_humans_2023, ragot_ai-generated_2020} – however, an AI-generated portrait of Alan Turing just sold for \$1.08 million \citep{cain_first_2024}, suggesting the opposite. On top of that, we propose another kind: the fusion of human and generative AI competencies to new levels of achievement. As AI’s capabilities continue to grow, its involvement in creative endeavors is set for further expansion and diversification. The examples from mathematics demonstrate that AI is no longer merely a tool but a collaborator in generating novel solutions. Moving forward, the challenge will be to strike the right balance: leveraging AI’s immense potential without undermining the unique contributions of human creativity, ensuring that the synergy between human intuition and AI’s capabilities leads to unprecedented creative achievements. Realizing this equilibrium is essential to ensure that AI is a complement and enhancer of human creativity rather than a substitute. Unlike traditional CSS, which facilitates the creative process primarily through knowledge processing and communication, generative AI systems possess the unique capacity to generate creative output independently. This marks a proactive step in the co-creative process, suggesting that AI can contribute in previously unimaginable ways.

However, this potential comes with challenges. A central question that mirrors debates about intelligence concerns the system boundaries we draw around creativity. Just as we ask, ``What is intelligent?'' we must also ask, ``What is creative?''. Is it the human using the tools, the tools themselves, or the synergetic combination of both? This question is critical because it determines how we assess the creativity of outputs in human-AI collaboration. If creativity is seen as emerging solely from the human, then AI's role is merely supportive. If, however, creativity is understood as a product of the combined efforts of humans and AI, then the co-creative process must be evaluated on its own terms, acknowledging the unique contributions of each entity. As humans use co-creative agents more intensely for their creative work, the risk of over-reliance on AI should not be overlooked. While AI can generate novel ideas and solutions that may not emerge from human creativity alone, there is a danger that excessive dependence on AI could undermine the unique aspects of human creativity, such as emotional depth, moral reasoning, and contextual awareness. This potential over-reliance emphasizes the importance of designing AI systems that support and amplify human creativity rather than diminish it.

In conclusion, integrating AI into creative work comes with scaling opportunities that are unheard of for creative advancements. The future of \hacc will hinge on balancing the enhancement, rather than substitution, of human creativity. Moving forward, the development of AI systems should focus on fostering collaboration rather than competition, enabling a harmonious fusion of human and machine creativity that pushes the boundaries of what is creatively possible. The concrete examples from the math field show us what is already possible in concise domains. Following the logic of the growth of generative AI tools in terms of efficiency, competencies, and generalizability, such co-creative efforts are expected to be possible in other domains soon.

\subsection*{Acknowledgments}

Jennifer Haase's work was supported by the German Federal Ministry of Education and Research (BMBF), grant number 16DII133 (Weizenbaum-Institute). Part of this work was conducted while Sebastian Pokutta was visiting Tokyo University via a JSPS International Research Fellowship. The authors would like to thank Christoph Spiegel for providing images of the colorings and Thomas Grisold for helpful comments on an early draft which significantly improved the exposition.   

\bibliographystyle{apalike}
\bibliography{references}

\end{document}